\documentstyle[epsf,aas2pp4]{article}
\newcommand{\SN}{SN1987A}

\newcommand{\freq}{843\,MHz}
\newcommand{\Aanda}[2]{{\aap},~{#1},~{#2}}
\newcommand{\Araa}[2]{{\araa},~{#1},~#2}
\newcommand{\Ajp}[2]{{Australian J. Phys.},~{#1},~{#2}}
\newcommand{\Apj}[2]{{\apj},~{#1},~{#2}}
\newcommand{\Nat}[2]{{Nature},~{#1},~{#2}}
\newcommand{\Pasa}[2]{{Proc.\ Astron.\ Soc.\ Australia},~{#1},~{#2}}
\newcommand{\sub}{6 June 2000}
\newcommand{\rev}{14 July 2000; 30 October 2000}
\newcommand{\acc}{31 October 2000}
\begin{document}
\begin{titlepage}
\vspace*{1.5cm}
\begin{center}
{\huge\bf RADIO SUPERNOVA 1987A AT 843 MHz}\\[2\baselineskip]
{\LARGE Lewis Ball$^1$, D. F. Crawford$^1$,
R. W. Hunstead$^1$, I. Klamer$^1$,\\[0.5\baselineskip]
\& V. J. McIntyre$^{1,2}$}\\[2\baselineskip]
{\large\em $^1$School of Physics, University of Sydney, N.S.W. 2006, Australia;
ball@physics.usyd.edu.au\\[\baselineskip]
$^2$ Australia Telescope National Facility, CSIRO, PO Box 76,
Epping, N.S.W. 2121, Australia}
\end{center}
\vspace*{\fill}
{\large 
{\bf The Astrophysical Journal}; in press\\
Submitted: \sub\\
Revised: \rev\\
Accepted: \acc
}
\end{titlepage}
\title{Radio Supernova 1987A at 843 MHz}
\author{Lewis Ball$^1$, D. F. Crawford$^1$, R. W. Hunstead$^1$,
I. Klamer$^1$, V. J. McIntyre$^{1,2}$}
\affil{$^1$ School of Physics, University of Sydney, N.S.W. 2006, Australia; ball@physics.usyd.edu.au}
\affil{$^2$ Australia Telescope National Facility, CSIRO, PO Box 76,
Epping, N.S.W. 2121, Australia}

\centerline{Astrophysical Journal, in press}
\centerline{Submitted: \sub . Revised: \rev . Accepted: \acc}


\begin{abstract}

We report here the flux densities of the evolving
radio source \SN\ at \freq\ measured from observations made
with the Molonglo Observatory Synthesis Telescope between
1994 September and 2000 May.
The radio light curve shows that the rate of increase of
the flux density jumped markedly around days 2800-3000
(i.e.\ in the first half of 1995),
and that since then the radio evolution has been remarkably well fitted
by a simple linear increase of $(62.7\pm 0.5)\, \mu{\rm Jy \, day^{-1}}$.
We discuss in detail the relationship between the radio light curve
and the recent brightening of the system at optical wavelengths.

\end{abstract}

\section{Introduction}

Supernova 1987A (\SN ) in the Large Magellanic Cloud is the nearest
supernova explosion detected in over 300 years.
Although not particularly bright, its proximity has facilitated studies
at a level of detail unprecedented in supernova research,
and rivalled only by the bright SN1993J.

We report here on observations of the radio emission from \SN\ at \freq\
using the Molonglo Observatory Synthesis Telescope (MOST).
Our results cover the period 1994 September -- 2000 May,
extending previously published data for
1987 February -- 1994 September (Ball et al.\ 1995).
Together these data sets from the MOST comprise a unique
time series, sampled at intervals of 2 to 6 weeks.

Radio emission from SN1987A was first detected with the MOST
on 1987 February 25.23 UT (Turtle et al.\ 1987),
just two days after the arrival of neutrinos on February 23.32
(Bionta et al.\ 1987).
After rising quickly to a peak flux density of $\sim 130$\,mJy at \freq\
the emission decayed rapidly and was undetectable at radio frequencies
after 1987 September.
On 1990 July 5 MOST observations indicated the reappearance of
radio emission from \SN\ (Ball et al.\ 1995),
an event confirmed about a month later by observations
at 4.8\,GHz with the Australia Telescope Compact Array
(Staveley-Smith et al.\ 1992).
These observations were the first indications that the 
expanding shock from the supernova explosion had encountered
a dramatic change in the circumstellar material (CSM)
it was passing through,
an interpretation first suggested on the basis of modelling
of the radio emission (Ball \& Kirk 1992).
Confirmation of this scenario followed from measurement of
the low expansion speed of the radio source
(Staveley-Smith et al.\ 1993; Gaensler at al.\ 1997)
which required a dramatic deceleration of the supernova shock
at or prior to the onset of the second phase of radio emission.
More detailed models for the radio emission (Duffy, Ball \& Kirk 1995)
and for the hydrodynamics of the interaction of the supernova ejecta
with the CSM
(e.g.\ Chevalier \& Dwarkadas 1995; Borkowski, Blondin \& McCray 1997)
all attribute the radio emission to the interaction of the
SN shock with a density jump in the CSM.

The implied radius of the circumstellar density jump responsible for
the reappearance of the radio emission is smaller than that of the
dense ring surrounding \SN\
(Jakobsen et al.\ 1991; Plait et al.\ 1995; Sonneborn et al.\ 1997).
Spatially-resolved spectra of \SN\ from the STIS instrument on the
Hubble Space Telescope indicated that on 1997 April 26 (day number 3714)
the reverse supernova shock radius was $\sim 80 \%$ of
that of the ring (Sonneborn et al.\ 1998).
HST images from the WFPC2 instrument show that
line emission from a local `knot' on the inner edge of the ring
brightened by tens of percent between early-1996 and mid-1997
(days 3200--3750; Garnavich, Kirshner \& Challis 1997).
There are now clear indications that similar brightenings are
taking place over a large fraction of the ring
(Bouchet et al.\ 2000; Lawrence \& Crotts 2000;
Maran, Pun \& Sonneborn 2000; Garnavich, Kirshner \& Challis 2000),
suggesting that the interaction of the supernova shock and the inner edge
of the ring is underway.
Further analysis of archival WFPC2 images shows that the optical signatures
of the interaction between material from the supernova explosion
and the inner edges of the ring have been detectable since early 1995
(Lawrence et al.\ 2000).

The radio data we report here show that the emission from \SN\
at \freq\ has continued to increase monotonically from 1990 July
until May 2000.
We present 106 new flux density determinations for
the radio emission from \SN\ at \freq\ between
1994 September and 2000 May.
In \S 2 we briefly describe the observations themselves.
In \S 3 (and Appendix A) we explain the approach used to reduce the
observations of this time-varying source.
The reduced data are described in \S 4 and tabulated in Appendix B.
In Appendix C we discuss an alternative,
more direct data reduction method,
and compare the results with those given in Appendix B.
In \S 5 we discuss the implications of the observed evolution
of the 843 MHz radio flux density from \SN.
In particular we investigate the evidence for features
in the 843 MHz radio light curve that may be related to the
dramatic changes in the optical line emission from this source
that herald the collision of the supernova shock with the ring.

\section{Observations}

The MOST is a synthesis telescope comprising two colinear,
cylindrical-paraboloid reflectors each $11.6\,$m wide and $778\,$m long
({Mills} 1981; {Robertson} 1991).
It operates at \freq\ (a wavelength of $0.356\,$m)
and measures right circular polarization.
The telescope is unique in that it forms
a comb of 64 fan beams in real time,
and an image is available
immediately following a 12-hour observation.
The synthesised beam of the telescope at the
declination of \SN\ is $43^{\prime\prime}\times 46^{\prime\prime}$ FWHM,
so the supernova is recorded as a point source.

We report here on the results of 106 12-hour synthesis
observations of the field including \SN.
All the observations reported here were made with the same pointing center,
using the standard $23^\prime$ field of the telescope.
A small number of observations which were incomplete,
or of low quality due to technical problems,
have been discarded.

\begin{figure}[htb]
%
%
\epsfxsize=8cm
\epsffile{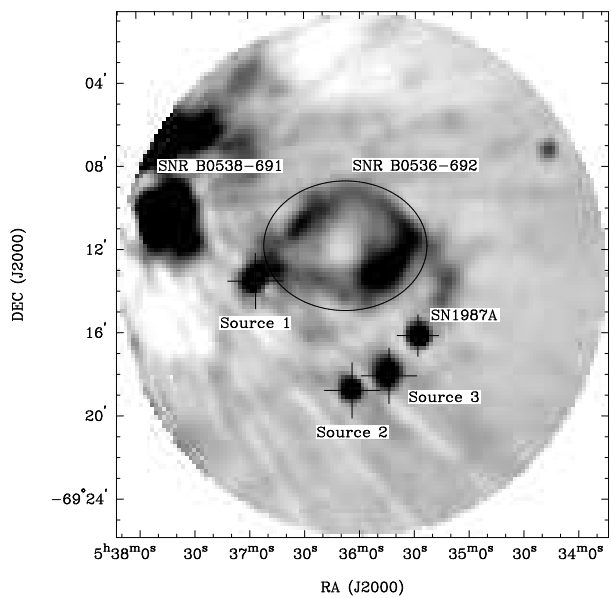}
\caption{
MOST grey-scale image of the \freq\ radio emission
on 2000 January 13 from the field including \SN.
Source 1, Source 2 and Source 3 are background objects
which were used for calibrating the flux density of the supernova.
The strong source on the eastern edge of the field is
SNR~B0538$-$691 and the linear artifacts that seem to originate
from it are in fact due to the very bright
H{\sc ii} region 30 Doradus which is outside the field.
The large ring of emission denoted 
SNR~B0536$-$692 may be from an old supernova remnant
but is now thought more likely to be a supershell.
The greyscale range runs from $-20\;\rm mJy/beam$ (white)
to $+50\;\rm mJy/beam$ (black).
\label{image}}
\end{figure}

\section{Data analysis}

The flux density of any unresolved source,
such as \SN, can be estimated directly from the image
produced automatically at the end of a 12-hour MOST synthesis observation.
However, the precision of such direct estimates is
limited by contributions from out-of-field sources which
cancel imperfectly as a result of variations in ambient conditions
and telescope characteristics during the observation.

\SN\ is situated in a complex region of the radio sky,
as can be seen from the MOST image of the field presented
in Figure \ref{image}.
The complexity of the radio emission surrounding \SN\
and, in particular, its proximity to the bright H{\sc ii}
complex 30 Doradus and the supernova remnant
SNR\,B0538$-$691, degrade the accuracy of direct estimates
of the flux density of \SN .
This was especially true in the first few years of observations
when the supernova was a relatively weak source.
As a result a novel
and complex method of analysing the data was developed specifically for
this unique object (Ball et al.\ 1995)
involving correcting each 24-second record of a 12-hour observation.
Although less transparent than procedures that aim to correct
the synthesised image, correction of each record in the data set
can provide flux density estimates with considerably lower scatter,
approaching the limiting thermal noise of the telescope. 

While the same method still provides accurate estimates of the flux
density of \SN, the effects of differential precession
(with respect to a reference data set made from observations before
the appearance of the supernova) have become important.
The increasing flux density of the supernova itself was also beginning
to affect the calibration procedure.
To address these effects some details of the analysis procedure
have been modified while retaining the significant features
of the method, particularly the reliance on correcting each 24-second
record in an observation to be as close as possible to that from the
reference.
The changes to the analysis from that of Ball et al.\ (1995)
are detailed in Appendix A.

The consistency between the flux density estimates obtained
by the revised data analysis described in Appendix A
and those published in Ball et al.\ (1995), is very good.
For the observations of the standard $23^\prime$ field up to
1994 September 3, 60\% of the estimates agree within $1\sigma$,
90\% within $2\sigma$ and 97\% within $3\sigma$,
where $\sigma$ is the uncertainty quoted by Ball et al.\ (1995) in
their Equation (2).

The internal consistency of the flux density measurements
can be estimated by analysing the scatter about a linear trend
for periods of up to a few hundred days over which such
a trend is a good fit to the data.
This suggests that the uncertainty at epochs after 1994 September
is dominated by a flux dependent term which is conservatively
estimated at 2\%, the same as the flux-dependent
contribution reported by Ball et al. (1995).

\section{Flux density estimates}

A total of 106 independent measurements of the \freq\
flux density from \SN\ are tabulated in Appendix B.
These comprise the MOST data set from 1994 September 17
(day 2763)
until 2000 May 05 (day 4820).
The quoted uncertainties for the record-corrected estimates
are simply 2\%, based on the scatter of the estimates about
a linear increase.

\begin{figure}[htb]
%
%
\epsfxsize=8.2cm
\epsffile{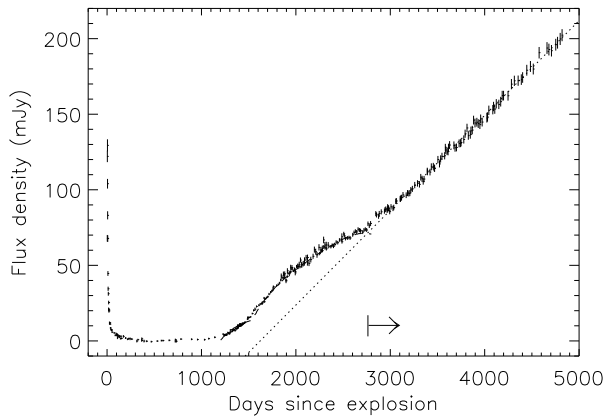}
\caption{Flux densities for \SN\ at \freq\
between 1987 February and 2000 May.
The arrow indicates the start of the new measurements
presented in this paper.
The dotted line shows a simple linear fit to the data at
epochs after day 3000.
The partly-obscured dashed line which extends to day 2800
shows the DBK95 model fit to the data from day 1000 to 2650.
\label{light}}
\end{figure}

The entire evolution of the radio emission from \SN\ at \freq,
based on the flux densities tabulated in Appendix B
together with those from Ball et al.\ (1995),
is shown in Figure \ref{light}.
This `light curve' illustrates dramatically the two distinct phases
of radio emission, separated by almost 3 years during which \SN\
was undetected at radio frequencies.

The most striking feature of the second phase of radio emission
is its continued steady increase over an interval of 10 years,
quite unlike the behaviour of
any other radio supernova observed to date
(e.g.\ Chevalier 1998 and references therein).
Some structure is visible in the light curve,
including a distinct jump in the rate of increase around days
2800--3000 which has not previously been reported.
A simple linear increase
at a rate of $(62.7\pm 0.5)\, \mu{\rm Jy \, day^{-1}}$ fits the
observed flux densities from day 3000 onwards remarkably well.
This rate of increase is significantly higher than that of 
$(31.8\pm1.4)\, \mu{\rm Jy \, day^{-1}}$
observed between days 2100 and 2700,
but not quite as high as the $(70.9\pm2.3)\, \mu{\rm Jy \, day^{-1}}$
increase seen between days 1500 and 2100 (Ball et al.\ 1995).
The \freq\ light curve shows
no evidence of any departure from the linear increase
between day 3000 (1995 May) and
4820 (2000 May).

\subsection{Template image subtraction}

The need for a complex treatment of the observations
of \SN\ has diminished as the supernova has increased in strength.
A simpler alternative method based on the subtraction of a
template image representing the observed field in the absence of
emission from \SN\ now provides reliable flux density estimates,
albeit with somewhat higher uncertainties.
This simplified analysis uses the standard data pipeline developed
for the Sydney University Molonglo Sky Survey
(SUMSS; Bock et al.\ 1999), and a comparison of
the technique with the more complicated  record-correction procedures
developed for \SN\ provides an invaluable test of its reliability.
Details of the template image subtraction estimates of the
flux densities of \SN\ for 48 observations between
1997 January 1 and 2000 May 05 are presented in
Appendix C.

\section{Discussion}

\subsection{Up to 1995}

The reappearance of radio emission from \SN\ in 1990 July
was first revealed by MOST observations at \freq ,
and is now attributed to the encounter of the expanding supernova shock
with a circumstellar density jump well inside the circumstellar ring
(Ball \& Kirk 1992 -- hereafter BK92;
Chevalier 1992;
Duffy, Ball \& Kirk 1995 -- hereafter DBK95). 
This interpretation is supported by the
analysis of images of the radio emission from \SN\
made at $8.6\,\rm GHz$ using the Australia Telescope Compact Array
(Staveley-Smith et al.\ 1993; Gaensler et al.\ 1997).
In particular, model fits to super-resolved images of the radio emission
suggest that the source of the radio emission was expanding at only
$\sim 2800\,\rm km\,s^{-1}$ in the period 1992--1995,
much lower than the average speed of $35,000\,\rm km\,s^{-1}$ required
between the time of explosion and the first imaging observations in 1992
(Gaensler et al.\ 1997).
While this expansion speed is not a direct measurement of the
speed of the supernova shock,
it clearly indicates that the shock had
undergone a dramatic deceleration from its initial speed of
greater than $50,000\,\rm km\,s^{-1}$
(Hanuschik \& Dachs 1987),
as would occur if the shock encountered a significant jump
in the density of circumstellar material
(Chevalier \& Dwarkadas 1995; Borkowski, Blondin \& McCray 1997)
associated with a dense H{\sc ii} region inside the circumstellar ring.
The interpretation is also consistent with the observed soft X-ray emission
from \SN\ (Beuermann, Brandt \& Pietsch 1994; 
Gorenstein, Hughes \& Tucker 1994;
Hasinger, Aschenbach \& Tr\"umper 1996),
though the X-ray evolution is relatively poorly sampled.

The only attempts to model the evolving second phase of radio emission
from \SN\ are those of BK92 and DBK95.
Based on diffusive acceleration of electrons at an
evolving supernova shock, including modification of the shock
hydrodynamics by the pressure of accelerated protons,
their model fits very well from switch-on to around day 2600
(as shown in Figure \ref{light}).
BK92 and DBK95 proposed that the
switch on and subsequent rise of the radio emission resulted from the
supernova shock encounter with a significant jump in the
density of the circumstellar material.
Chevalier \& Dwarkadas (1995) placed this suggestion on a firmer footing,
indicating that a very large density jump was
expected at the inner boundary of a dense H{\sc ii} region
surrounding the progenitor.
At times later than those considered
by DBK95 (i.e. after about day 2600 or May 1995) the model flux
density reaches a plateau and then decreases slowly
-- contrary to the observations.
The DBK95 model has many limitations (such as assuming spherical
symmetry) which rapidly become less tenable at such epochs.
No detailed model that combines a plausible distribution of
the circumstellar material with a calculation of the supernova shock
hydrodynamics and which treats the particle acceleration self
consistently has yet been developed.

\subsection{1995 on}

The optical brightening of the inner edge of the circumstellar ring
that has been observed from mid-1996 on 
(Pun et al.\ 1997; Garnavich, Kirshner \& Challis 1997, 1999 \& 2000;
Bouchet et al.\ 2000; Lawrence \& Crotts 2000;
Maran, Pun \& Sonneborn 2000)
clearly indicates that the interaction of the supernova shock
and the ring is underway.
H$\alpha$ images show that an isolated spot just inside the
main ring brightened by roughly 50\% between
1994 February (day $\sim 2500$)
and 1997 July (day $\sim 3750$)
and broad band (R) images indicate a brightening
by $\sim 33$\%, occurring between 1996 February (day $\sim 3300$)
and 1997 July (Garnavich, Kirshner \& Challis 1997).
More recent reports of R-band observations state that the
brightening over the four years prior to 1999 February are well
fitted by a steady increase in brightness,
suggesting that the optical signature of the interaction
may have begun as early as 1995 February
(day $\sim 2900$; Garnavich, Kirshner \& Challis 1999).
This has been confirmed by the analysis of archival HST data
by Lawrence et al.\ (2000).

It is possible that the jump in the
rate of increase of the radio emission around days 2800 -- 3000
and the subsequent long-lived linear rise
is associated with the impact of the supernova shock
with the ring material.
If not, the lack of features in the \freq\
light curve between days 2800 and 4820 implies that there
is no significant radio emission
from the impact sites evident in the optical images.
Unfortunately, the highest resolution presently available at radio
wavelengths with sufficient sensitivity is $\sim 0.4$ arcsec
at 9 GHz (Gaensler et al.\ 1997) so direct comparison of radio
and optical images is not definitive in this regard.

\subsubsection{Timing and the shock radius}

The optical `hot spots' which were the first indications of the
shock-ring interaction have been reported as being at position angle
$\sim 30^\circ$ and $0.55\;$arcsec from the supernova
($87\pm 4$\% of the ring radius; Garnavich, Kirshner \& Challis 1997)
and at p.a.\ $103^\circ\pm5^\circ$ and $0.6-0.65\;$arcsec 
(around 80\% of the ring radius;
Bouchet et al.\ 2000; Lawrence \& Crotts 2000).
Lawrence et al.\ (2000) have shown that there was detectable
optical emission from the first hot spot (near p.a.\ $30^\circ$)
since at least March 1995 (day $\sim 2950$).
This first optical hot spot is not coincident with
either radio lobe seen in the $9\;$GHz images of
Gaensler et al.\ (1997).
The brightening first seen around p.a.\ $100^\circ$ has now been
shown to extend over position angles from $90^\circ$ to $130^\circ$
(Maran, Pun \& Sonneborn 2000), which is roughly coincident with the
brighter of the two radio `lobes' or `hot spots' seen in the
$9\;$GHz images of Gaensler et al.\ (1997) --
and with the major axis of the ring.
Lawrence et al.\ (2000) have shown that the emission from this region
began no later than 1999 January (day $\sim 4350$)
and probably as early as 1998 February (day $\sim 4000$).

If the first optical hot spot is in the ring plane,
its projected radial distance of $\sim 0.55\;$arcsec and
p.a.\ of $30^\circ$
imply a true radial distance from the progenitor of
$\sim 0.7\;$arcsec.
The appearance of this emission in early 1995
(around day 3000) therefore suggests a shock radius which
is consistent with or perhaps slightly larger than that
inferred from model fits of a spherical source to the radio images
(Gaensler et al.\ 1997, Figure 10),
given the significant uncertainties in both estimates.
The near coincidence of the second more extended
region of optical emission with the major axis of the ring implies
that if the radiating material is in the ring plane,
its true radial distance from the progenitor is the same as its
projected radial distance in the plane of the sky. 
The appearance of this optical emission at a radial distance of
$\sim 0.65\;$arcsec around the beginning of 1998 is again consistent
with the radius of model fits to the radio images.
Furthermore, the very similar radial distances to the first
and second optical hot spots,
and the significant delay between their appearance,
are broadly consistent with the low expansion speed
inferred from the fits to the radio images
(Gaensler et al.\ 1997).

\subsubsection{Acceleration times}

The acceleration time required for electrons to attain
sufficient energy to produce radio frequency synchrotron emission
is an important physical parameter.
The appearance of optical hot spots suggest that the
shock impact with the ring began at least as early as March 1995
(day $\sim 2950$; Lawrence et al.\ 2000).
If the change in the evolution of the \freq\ radio emission around
day 3000 is related to the ring impact,
the implied acceleration time is less than a few hundred days.
The alternative, that a change in the radio
emission associated with the ring impact is yet to be observed,
may indicate an acceleration time greater than
$\sim 1800$ days, the interval from 1995 February to 2000 May.
The model for the radio reappearance of \SN\
in 1990 July (BK92; DBK95) suggested that
the acceleration time was $\sim 300\;\rm days$.

Theory suggests that the time required for diffusive shock acceleration
scales as $(B v_s^2)^{-1}$ where $B$ is the magnetic
field and $v_s$ is the shock speed.
Given that the shock speed is most probably decreasing with time,
the first scenario suggests a significant increase in $B$ between
the first circumstellar density jump and the ring,
while the latter suggests a decrease.
If the field is that of the stellar wind,
an increase in $B$ with radial distance occurs
if the wind is confined (as it is in \SN),
but more detailed modelling is required to address
such issues adequately.

\section{Future outlook}

The very good fit of a simple linear increase to the evolving \freq\
emission from early 1995 to mid-2000 is remarkable given the
inherent complexity of this source.
Modelling of this dependence is beyond the scope of the present paper,
but its simplicity is encouraging.
We propose that the transition to the linear increase
at \freq\ in early 1995 is related to the impact of
the supernova shock with circumstellar ring material
seen at optical wavelengths by the HST,
or at least with fingers of very dense material which intrude from the
inner edge of the ring.
On the other hand, if the onset of radio emission associated
with the ring impact has not yet occurred,
it is likely that the first detection of this event will
result from the ongoing regular monitoring of \SN\ with the MOST.
Comparisons of the \freq\ light curves with
Australia Telescope Compact Array data --
comprising light curves at higher radio frequencies and images of the
$9\;$GHz emission -- are underway.
Analysis aimed at identifying possible changes in the
radio spectral index are also underway.

There is ample evidence from observations of the CSM,
the explosion ejecta, images of the radio emission
and now the brightening of the ring material
that the \SN\ system is now far from spherically symmetric.
The relationship between the radio emission,
the optical emission and the shock evolution is likely to carry
strong signatures of this geometry.
It is also possible that the radio and optical emission originate
from distinct regions of the shock which are encountering different
circumstellar environments, and are at different radii.
The imminent upgrade of the Australia Telescope Compact Array
to 25 GHz will provide higher angular resolution radio images;
and comparison with optical and new X-ray images from
Chandra may help to clarify such issues.

\section*{Acknowledgements}

The authors thank Duncan Campbell-Wilson, the Site Manager at the
Molonglo Radio Observatory, for his continued efforts and enthusiasm in
pursuing the monitoring program for \SN .
We also thank Bryan Gaensler for numerous ongoing
productive and stimulating discussions.
The MOST is operated by the University of Sydney and supported 
in part by grants from the Australian Research Council.

\appendix

\vspace*{-\baselineskip}
\section{Data reduction}

Each 12-hour MOST synthesis observation of a $23^\prime$ field
consists of approximately 1795 24-second records comprising
128 beam responses.
The analysis used by Ball et al.\ (1995) involved correcting each
record of an observation to be as close as possible to that from
a reference data set generated by combining 50 observations
of the same field.
The analysis used to calculate the flux densities presented here
is very similar but with the following differences.

The flux density of \SN\ in a given observation was first estimated
directly from the observed data by fitting a point source
at the appropriate location to each record.
A point source with the mean flux density determined by this method
was added to the reference data set at the position of \SN.
The observed data were then interpolated to the hour angle of the
amended reference data set.
(In the earlier analysis of Ball et al.\ 1995 the reference data were
interpolated to the hour angle of each observation.)
Records with an unacceptably high noise level were
rejected at this stage.
A gain correction was then determined for each record in the
interpolated observation by performing a least squares fit of 
the beam responses in the data to those in the corresponding amended
reference record.

The amended reference data were then subtracted from the interpolated
observed data.
If the direct estimate of the supernova flux density was accurate,
the difference data should contain only noise.
We test that assumption by fitting a point source to each record of the
difference data, at the position of \SN .
If the assumption is correct, the mean flux density will be zero,
and the record-to-record rms variation will indicate the uncertainty.
In general the mean is small but not zero, indicating the effects
of imperfect cancellation of out of field sources due to
variations in telescope performance during the observation.
To correct for these effects we sum the direct estimate
of the flux density and the non-zero mean of the fits to the
difference data.

The flux density was then scaled by the factor required to ensure
that the sum of the flux densities of the
nearby reference sources (Sources 1, 2 \& 3) was constant.  
The final estimate of the flux density of \SN\ was obtained
by applying a fixed offset to correct for a non-zero base level
in the reference data at the supernova position.

In the analysis of Ball et al.\ (1995) correction terms for the
gain, a phase difference and a phase gradient were obtained
through a least-squares fit of the Fourier transform of each
24-second record to the corresponding transform of
the reference data.
The reference data were then subtracted from
the inverse transform of the corrected data,
and a difference image was synthesised.
The flux density of \SN\ was estimated
by fitting a source to the difference image.
In the new analysis, the comparison was made without performing the
Fourier transforms, and the fitting to the difference data
was performed for each record, not for the synthesised image.
The decision to avoid Fourier transforming the data
has the advantage that it avoids the scattering of spurious
signal across the data sets that sometimes occurs in the
inverse transform
(primarily when the untransformed data have much more power at
one end of the vector),
increasing the noise levels.

\vspace*{-\baselineskip}
\section{Flux densities}

The table contains the observed flux densities $S$ from
\SN\ at \freq\ from 1994 September 17 to 2000 May 05.
The day number refers to the midpoint of the observation in UT,
relative to the geocentric time of the explosion on
1987 February 23.32 (UT) or JD 2446849.82.
The uncertainty estimates $\sigma$
are 2\% of the flux density.

A complete file of the \freq\ flux densities of \SN\
is available from\\
http://www.physics.usyd.edu.au/astrop/SN1987A

\newpage

\centerline{
\begin{tabular}{lrrr}
\hline \hline
\multicolumn{1}{c}{Date} & 
\multicolumn{1}{c}{Day} & 
\multicolumn{1}{c}{$S$} & 
\multicolumn{1}{c}{$\sigma$} \\
\multicolumn{1}{c}{(UT)} &
\multicolumn{1}{c}{Number} &
\multicolumn{1}{c}{(mJy)} &
\multicolumn{1}{c}{(mJy)} \\
\hline
    1994 Sep 17 &    2763.5 &      77.5 &       1.5 \\
    1994 Oct 01 &    2777.5 &      76.2 &       1.5 \\
    1994 Oct 15 &    2791.4 &      77.7 &       1.6 \\
    1994 Dec 09 &    2846.3 &      84.2 &       1.7 \\
    1994 Dec 17 &    2854.3 &      83.4 &       1.7 \\
    1995 Jan 08 &    2876.2 &      82.5 &       1.7 \\
    1995 Jan 15 &    2883.2 &      84.3 &       1.7 \\
    1995 Feb 05 &    2904.1 &      85.4 &       1.7 \\
    1995 Mar 11 &    2938.0 &      85.9 &       1.7 \\
    1995 Mar 18 &    2945.0 &      86.9 &       1.7 \\
    1995 Mar 26 &    2953.0 &      87.3 &       1.7 \\
    1995 Apr 08 &    2966.0 &      86.4 &       1.7 \\
    1995 Apr 14 &    2971.9 &      88.7 &       1.8 \\
    1995 Apr 16 &    2973.9 &      86.8 &       1.7 \\
    1995 Apr 22 &    2979.9 &      88.0 &       1.8 \\
    1995 Apr 25 &    2982.9 &      87.7 &       1.8 \\
    1995 May 06 &    2993.9 &      85.5 &       1.7 \\
    1995 May 20 &    3007.8 &      88.7 &       1.8 \\
    1995 Jun 10 &    3028.8 &      87.7 &       1.8 \\
    1995 Jun 18 &    3036.8 &      88.3 &       1.8 \\
    1995 Jul 09 &    3057.7 &      92.4 &       1.8 \\
    1995 Jul 16 &    3064.7 &      92.6 &       1.9 \\
    1995 Aug 25 &    3105.6 &      94.1 &       1.9 \\
    1995 Aug 26 &    3106.6 &      93.8 &       1.9 \\
    1995 Sep 02 &    3113.6 &      94.7 &       1.9 \\
    1995 Sep 08 &    3119.5 &      94.8 &       1.9 \\
    1995 Sep 23 &    3134.5 &      96.6 &       1.9 \\
    1995 Oct 27 &    3168.4 &      96.4 &       1.9 \\
    1995 Nov 10 &    3182.4 &      97.8 &       2.0 \\
    1995 Nov 24 &    3196.3 &      97.6 &       2.0 \\
    1995 Dec 05 &    3207.3 &      97.8 &       2.0 \\
    1995 Dec 15 &    3217.3 &      99.8 &       2.0 \\
    1996 Jan 01 &    3234.2 &     100.6 &       2.0 \\
    1996 Jan 26 &    3259.2 &     103.1 &       2.1 \\
    1996 Feb 24 &    3288.1 &     104.2 &       2.1 \\
    1996 Mar 16 &    3309.0 &     103.6 &       2.1 \\
    1996 Mar 30 &    3323.0 &     108.3 &       2.2 \\
    1996 Apr 13 &    3336.9 &     107.2 &       2.1 \\
    1996 Apr 25 &    3348.9 &     105.5 &       2.1 \\
    1996 May 04 &    3357.9 &     107.1 &       2.1 \\
    1996 Jun 07 &    3391.8 &     112.4 &       2.2 \\
    1996 Jun 08 &    3392.8 &     111.8 &       2.2 \\
\hline
\end{tabular}
}

\centerline{
\begin{tabular}{lrrr}
\hline \hline
\multicolumn{1}{c}{Date} & 
\multicolumn{1}{c}{Day No.} & 
\multicolumn{1}{c}{$S$} & 
\multicolumn{1}{c}{$\sigma$} \\
\multicolumn{1}{c}{(UT)} &
\multicolumn{1}{c}{Number} &
\multicolumn{1}{c}{(mJy)} &
\multicolumn{1}{c}{(mJy)} \\
\hline
    1996 Jul 06 &    3420.7 &     111.7 &       2.2 \\
    1996 Jul 19 &    3434.7 &     115.2 &       2.3 \\
    1996 Aug 09 &    3455.6 &     115.5 &       2.3 \\
    1996 Aug 17 &    3463.6 &     114.6 &       2.3 \\
    1996 Aug 23 &    3469.6 &     114.1 &       2.3 \\
    1996 Sep 27 &    3504.5 &     120.0 &       2.4 \\
    1996 Oct 14 &    3521.4 &     117.1 &       2.3 \\
    1996 Oct 26 &    3533.4 &     120.4 &       2.4 \\
    1996 Nov 09 &    3547.4 &     122.1 &       2.4 \\
    1996 Dec 08 &    3576.3 &     122.5 &       2.4 \\
    1996 Dec 15 &    3583.3 &     127.8 &       2.6 \\
    1996 Dec 19 &    3587.3 &     126.3 &       2.5 \\
    1997 Jan 02 &    3601.2 &     125.5 &       2.5 \\
    1997 Jan 13 &    3612.2 &     125.3 &       2.5 \\
    1997 Jan 17 &    3616.2 &     127.9 &       2.6 \\
    1997 Jan 29 &    3628.2 &     129.6 &       2.6 \\
    1997 Mar 16 &    3674.0 &     129.4 &       2.6 \\
    1997 Mar 28 &    3686.0 &     127.8 &       2.6 \\
    1997 Apr 12 &    3700.9 &     128.0 &       2.6 \\
    1997 May 10 &    3728.9 &     130.7 &       2.6 \\
    1997 May 24 &    3742.8 &     130.8 &       2.6 \\
    1997 May 31 &    3749.8 &     131.3 &       2.6 \\
    1997 Jun 29 &    3778.7 &     133.3 &       2.7 \\
    1997 Jul 13 &    3792.7 &     137.1 &       2.7 \\
    1997 Aug 01 &    3812.6 &     140.9 &       2.8 \\
    1997 Aug 15 &    3826.6 &     135.8 &       2.7 \\
    1997 Sep 05 &    3847.5 &     136.5 &       2.7 \\
    1997 Sep 12 &    3854.5 &     139.0 &       2.8 \\
    1997 Sep 26 &    3868.5 &     139.5 &       2.8 \\
    1997 Oct 10 &    3882.4 &     146.1 &       2.9 \\
    1997 Oct 25 &    3897.4 &     144.2 &       2.9 \\
    1997 Nov 14 &    3917.3 &     144.7 &       2.9 \\
    1997 Nov 28 &    3931.3 &     143.8 &       2.9 \\
    1997 Dec 16 &    3949.3 &     143.1 &       2.9 \\
    1997 Dec 24 &    3957.2 &     146.9 &       2.9 \\
    1998 Jan 09 &    3973.2 &     144.8 &       2.9 \\
    1998 Feb 28 &    4023.1 &     147.5 &       3.0 \\
    1998 Mar 07 &    4030.0 &     148.1 &       3.0 \\
    1998 Mar 21 &    4044.0 &     152.6 &       3.1 \\
    1998 May 02 &    4085.9 &     153.0 &       3.1 \\
    1998 May 10 &    4093.9 &     155.1 &       3.1 \\
    1998 May 23 &    4106.8 &     153.4 &       3.1 \\
    1998 Jun 08 &    4122.8 &     156.9 &       3.1 \\
\hline
\end{tabular}
}

\centerline{
\begin{tabular}{lrrr}
\hline \hline
\multicolumn{1}{c}{Date} & 
\multicolumn{1}{c}{Day No.} & 
\multicolumn{1}{c}{$S$} & 
\multicolumn{1}{c}{$\sigma$} \\
\multicolumn{1}{c}{(UT)} &
\multicolumn{1}{c}{Number} &
\multicolumn{1}{c}{(mJy)} &
\multicolumn{1}{c}{(mJy)} \\
\hline
    1998 Jul 05 &    4149.7 &     156.5 &       3.1 \\
    1998 Jul 24 &    4169.7 &     158.3 &       3.2 \\
    1998 Aug 09 &    4185.6 &     162.3 &       3.2 \\
    1998 Aug 23 &    4199.6 &     162.3 &       3.2 \\
    1998 Oct 10 &    4247.4 &     162.2 &       3.2 \\
    1998 Nov 15 &    4283.3 &     169.9 &       3.4 \\
    1998 Dec 16 &    4314.3 &     172.1 &       3.4 \\
    1999 Jan 25 &    4354.2 &     172.2 &       3.4 \\
    1999 Feb 28 &    4388.1 &     172.9 &       3.5 \\
    1999 Mar 14 &    4402.0 &     174.3 &       3.5 \\
    1999 Apr 27 &    4445.9 &     179.1 &       3.6 \\
    1999 Jun 05 &    4484.8 &     181.6 &       3.6 \\
    1999 Jul 05 &    4514.7 &     179.9 &       3.6 \\
    1999 Sep 04 &    4576.5 &     190.7 &       3.8 \\
    1999 Nov 27 &    4660.3 &     193.8 &       3.9 \\
    1999 Dec 17 &    4680.3 &     192.8 &       3.9 \\
    2000 Jan 13 &    4707.2 &     192.0 &       3.8 \\
    2000 Feb 28 &    4753.1 &     196.0 &       3.9 \\
    2000 Mar 18 &    4772.0 &     197.3 &       3.9 \\
    2000 Apr 15 &    4799.9 &     199.8 &       4.0 \\
    2000 May 05 &    4819.9 &     202.3 &       4.0 \\
\hline
\end{tabular}
}

\vspace*{-0.75\baselineskip}
\section{Template image subtraction}

This simplified analysis uses a template image constructed from
observations performed in the period between the prompt and
second phases of radio emission from \SN. 
Four full 12-hour synthesis observations (on 1989 Feb 15, 1989 Apr 15,
1989 Dec 20 and 1990 Apr 08) were selected,
scaled and combined to produce the template.
A set of scaled versions of the template image was then constructed,
stepping the scaling factor from 0.9 to 1.1 in increments of 0.02.

The object denoted in Figure \ref{image} as `Source 2'
-- the only point source close to \SN\ and reasonably free of
background confusion --
is of primary importance in the calibration.
The scaling of the template was such that the flux density
determined by a point source fit to Source 2 had a 
flux density of $133.0\,\rm mJy$.
The other background objects denoted `Source 1' and `Source 3'
were also used for calibration but were given less weight;
Source 1 sits on a radial artifact from 30 Doradus
and Source 3 is marginally extended.

\begin{figure}[htb]
%
%
\epsfxsize=8cm
\epsffile{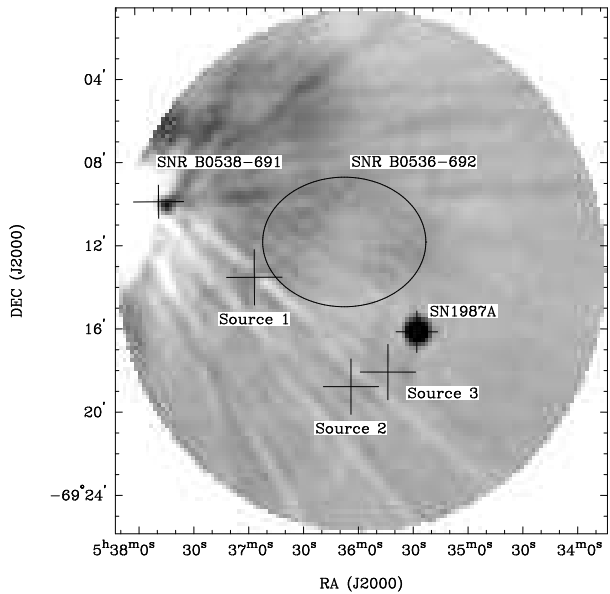}
\caption{Grey-scale image of the difference between the
original image shown in Figure \ref{image} and the scaled template
which minimises the flux density at the peak pixel of Source 2.
The greyscale range runs from $-20\;\rm mJy/beam$ (white)
to $+50\;\rm mJy/beam$ (black).
The strong signal at the position
of \SN\ indicates the increase in the SN flux density from early
1990 to 2000 January. 
The strong negative source seen as a white region on the
eastern edge of the image is near, but not coincident with,
the peak of SNR~B0538$-$691.
It indicates a source which has faded
considerably since 1990.
\label{difference}}
\end{figure}

The simplified process of measuring the flux density from a
given observation involves the following steps:
\begin{enumerate}
\item
A set of difference images is constructed by subtracting
the scaled templates from the observed image.
\item
The pair of difference images in which the flux densities in the
peak pixel of the chosen reference source (Source 1, 2 or 3)
are closest to zero (one negative and one positive)
is identified.
One such difference image is presented in Figure \ref{difference}.
The scaling factor for which the peak flux density of the reference source
would be zero is then determined by simple linear interpolation.
\item
Step 2 is repeated for each of the other two reference sources.
\item
The three estimates of the scaling factor are averaged,
with the estimate based on Source 2 given double the weight of the
estimates based on Sources 1 and 3.
\item
The nominal peak flux density of the source at the position
of \SN\ is determined by fitting a Gaussian profile to the relevant
region of the previously-chosen difference images.
\item
The nominal flux density is divided by the appropriate mean
scale factor.
\end{enumerate}

The scale factors determined by this process show an RMS scatter
of 4.3\% about unity which is consistent with the calibration uncertainty
of the Molonglo telescope.
We estimate the total random uncertainty in the flux density estimates
obtained by template image subtraction to be $\sim 3$\%.

\begin{figure}[htb]
%
%
\epsfxsize=8.2cm
\epsffile{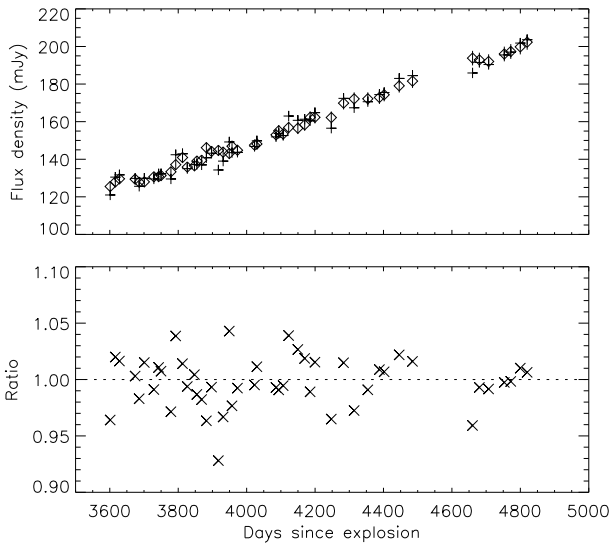}
\caption{Flux densities for \SN\ at \freq\
between 1997 January and 2000 May (top panel).
The diamonds show the estimates calculated from the
record-correction technique as presented in Appendix B
with error bars indicating the quoted 2\% uncertainties.
The $+$ symbols show the estimates from the same observations based
on the simpler template image subtraction method.
Error bars indicating the 3\% uncertainties in these estimates
have been omitted for clarity.
The lower panel shows the ratios of the two estimates.
\label{compare}}
\end{figure}

Figure \ref{compare} shows a comparison between
the two alternative estimates of the
flux density of \SN\ for 48 epochs between 1997 January
and 2000 May. 
The consistency between the two methods is very good.
The template image subtracted estimates have slightly larger
uncertainties than the record-corrected data,
as reflected in a larger scatter about the trend in Figure \ref{compare}.
The flux densities estimated by template image subtraction differ by at
most 6\% from those determined by record correction,
which is consistent with the combined uncertainties
in the two data sets.
The ratios of the two estimates are plotted
in the lower panel of Figure \ref{compare};
the mean ratio and its standard error is $0.998\pm 0.003$.
We conclude that there is no significant systematic difference between
the two methods of analysis when applied to post 1997 January
observations.

\clearpage


\end{document}